\documentclass[11pt,onecolumn,prl]{revtex4}
\usepackage{amssymb}
\usepackage{amsmath}
\usepackage{graphicx}
\usepackage{epsfig}
\usepackage{color}
\usepackage{makeidx}

\begin{document}
\title{Synthesis and physical properties of the new potassium iron
selenide superconductor K$_{0.80}$Fe$_{1.76}$Se$_{2}$}
\author{R. Hu,$^{1,2}$ E. D. Mun,$^{3}$ D. H. Ryan,$^{4}$ K. Cho,$^{2}$ H.
Kim,$^{1,2}$ H. Hodovanets,$^{1,2}$ W. E. Straszheim,$^{2}$ M. A. Tanatar,$%
^{2}$ R. Prozorov$^{1,2}$, W. N. Rowan-Weetaluktuk,$^{4}$ J. M. Cadogan,$%
^{5} $ M. M. Altarawneh,$^{3}$ C. H. Mielke,$^{3}$ V. S. Zapf,$^{3}$ S. L.
Bud'ko,$^{1,2}$ P. C. Canfield$^{1,2}$}
\affiliation{$^{1}$Department of Physics and Astronomy, Iowa State University, Ames, IA 50011, USA}
\affiliation{$^{2}$Ames Laboratory, U.S. DOE, Ames, IA 50011, USA}
\affiliation{$^{3}$National High Magnetic Field Laboratory, Los Alamos National Laboratory, Los Alamos, New Mexico 87544, USA}
\affiliation{$^{4}$Physics Department and Centre for the Physics of Materials, McGill University, Montreal, H3A 2T8, Canada}
\affiliation{$^{5}$Department of Physics and Astronomy, University of Manitoba, Winnipeg, Manitoba, R3T 2N2, Canada}

\begin{abstract}
In this article we review our studies of the K$_{0.80}$Fe$_{1.76}$Se$_{2}$
superconductor, with an attempt to elucidate the crystal growth details and
basic physical properties over a wide range of temperatures and applied
magnetic field, including anisotropic magnetic and electrical transport
properties, thermodynamic, London penetration depth, magneto-optical imaging
and M\"{o}ssbauer measurements. We find that: (i) Single crystals of similar
stoichiometry can be grown both by furnace-cooled and decanted methods; (ii)
Single crystalline K$_{0.80}$Fe$_{1.76}$Se$_{2}$ shows moderate anisotropy
in both magnetic susceptibility and electrical resistivity and a small
modulation of stoichiometry of the crystal, which gives rise to broadened
transitions; (iii) The upper critical field, $H_{c2}$(T) is $\sim $55 T at 2
K for $\mathbf{H}\parallel \mathbf{c}$, manifesting a temperature dependent
anisotropy that peaks near 3.6 at 27 K and drops to 2.5 by 18 K; (iv) M\"{o}%
ssbauer measurements reveal that the iron sublattice in K$_{0.80}$Fe$_{1.76}$%
Se$_{2}$ clearly exhibits magnetic order, probably of the first order, from
well below $T_{c}$ to its N\'{e}el temperature of $T_{N}=\,532\,\pm \,2$~K.
It is very important to note that, although, at first glance there is an
apparent dilemma posed by these data: high $T_{c}$ superconductivity in a
near insulating, large ordered moment material, analysis indicates that the
sample may well consist of two phases with the minority superconducting
phase (that does not exhibit magnetic order) being finely distributed, but
connected with in an antiferromagnetic, poorly conducting, matrix,
essentially making a superconducting aerogel.
\end{abstract}
\maketitle

\section{Introduction}

The iron-based superconductors have attracted intense research attention
because of their high transition temperature and their possibly
unconventional pairing mechanism, correlated to magnetism.\cite{Kenji}$-$%
\cite{Johnpierre} Similar to cuprate superconductors, iron-based
superconductors have layered structures; the planar Fe layers tetrahedrally
coordinbated by As or chalcogen anions (Se or Te) are believed to be
responsible for superconductivity. Stacking of the FeAs building blocks with
alkali, alkaline earth or rare earth oxygen spacer layers forms the basic
classes of iron arsenic superconductors in these compounds: 111-type AFeAs%
\cite{Wang}, 122-type AFe$_{2}$As$_{2}$\cite{Rotter}$-$\cite{Jasper},
1111-type ROFeAs\cite{Kamihara}$,$\cite{Chen} and more complex block
containing phases, e.g. Sr$_{2}$VO$_{3}$FeAs\cite{Zhu}, Sr$_{3}$Sc$_{2}$Fe$%
_{2}$As$_{2}$O$_{5}$\cite{Zhu2}, Sr$_{4}$Sc$_{2}$Fe$_{2}$As$_{2}$O$_{6}$.%
\cite{Chen2} The simple binary 11-type iron chalcogenide has no spacer
layers and superconductivity can be induced by doping FeTe with S\cite%
{Rongwei} or Se.\cite{Mizu} Different from the other iron-based
superconductors, FeSe is a superconductor\cite{Hsu}, $T_{c}\sim 8$ K,\ with
no static magnetic order and its transition temperature can be increased up
to 37 K by applying pressure\cite{Med} or 15 K in FeSe$_{0.5}$Te$_{0.5}$.%
\cite{Mizu} More recently, superconductivity above 30 K has been reported in
A$_{x}$Fe$_{2-y}$Se$_{2}$ (A = K, Cs, Rb or Tl)\cite{Guo}$-$\cite{AFWang},
by adding A between the Fe$_{2}$Se$_{2}$ layers, a compound with the same
unit cell structure as the AFe$_{2}$As$_{2}$ compounds.

$\mu SR$ measurements showed that magnetic order co-exists with bulk
superconductivity in Cs$_{0.8}$Fe$_{1.6}$Se$_{2}$\cite{Sher}, and neutron
diffraction measurements on K$_{0.8}$Fe$_{1.76}$Se$_{2}$ \cite{Weibao} have
suggested that not only do magnetic order and superconductivity co-exist,
but that the iron moments are remarkably large (3.3~$\mu _{B}$/Fe) and are
ordered in a relatively complex antiferromagnetic structure that places all
of the iron moments parallel to the \textit{c}-axis. The magnetic ordering
temperatures are quite high in both compounds: $T_{N}$(Cs)=480~K \cite{Sher}%
, $T_{N}$(K)=560~K \cite{Weibao}. The development of a paramagnetic
component near $T_{N}$ \cite{Sher} and the unusual temperature dependence of
the magnetic intensity \cite{Weibao} suggest that the magnetic transition
may be first order in nature rather than being a more conventional second
order transition. First order magnetic transitions are commonly associated
with changes in crystal structure, and both synchrotron x-ray diffraction 
\cite{Pomj1919} and neutron diffraction \cite{Weibao}$,$\cite{Weibao2} have
now shown evidence for a structural change from $I4/m$ to $I4/mmm$
associated with a disordering of iron vacancies that occurs in the vicinity
of the magnetic transition. In this review we summarize our basic
understanding of this material.\cite{Hu1931}$-$\cite{Ryan2011} First we
clarify the growth details and present elemental analysis and physical
properties of K$_{0.80}$Fe$_{1.76}$Se$_{2}$ single crystals.\cite{Hu1931}
Then the $H_{c2}$-$T$ phase diagram for the K$_{0.8}$Fe$_{1.76}$Se$_{2}$ is
constructed and discussed.\cite{Mun} At the end we present the study of the
magnetic ordering of K$_{0.80}$Fe$_{1.76}$Se$_{2}$ using $^{57}$Fe M\"{o}%
ssbauer spectroscopy.\cite{Ryan2011}

\section{Experimental methods}

Crystals were characterized by powder x-ray diffraction using a Rigaku
Miniflex x-ray diffractometer. The actual chemical composition was
determined by wavelength dispersive x-ray spectroscopy (WDS) in a JEOL
JXA-8200 electron microscope. Magnetic susceptibility was measured in a
Quantum Design MPMS, SQUID magnetometer. In plane AC resistivity $\rho _{ab}$
was measured by a standard four-probe configuration. Measurement of $\rho
_{c}$ was made in the two-probe configuration. Contacts were made by using a
silver alloy. For $\rho _{c}$, contacts were covering the whole \textit{ab }%
plane area.\cite{Jeffrey} Thermoelectric power\ measurements were carried
out by a dc, alternating temperature gradient (two heaters and two
thermometers) technique.\cite{MunTEP} Specific heat data were collected
using a Quantum Design PPMS. The in-plane London penetration depth was
measured by using a tunnel-diode resonator (TDR) oscillating at 14 MHz and
at temperature down to 0.5 K.\cite{RuslanTDR} Magneto-optical imaging was
conducted by utilizing the Faraday effect in bismuth-doped iron garnet
indicators with in-plane magnetization.\cite{Doro} A flow-type liquid $^{4}$%
He cryostat with sample in vacuum was used. The sample was positioned on top
of a copper cold finger and an indicator was placed on top of the sample.
The cryostat was positioned under polarized-light reflection microscope and
the color images could be recorded on video and high-resolution CCD cameras.
When linearly polarized light passes through the indicator and reflects off
the mirror sputtered on its bottom, it picks up a double Faraday rotation
proportional to the magnetic field intensity at a given location on the
sample surface. Observed through the (almost) crossed analyzer, we recover a
2D image.\cite{Joos}

To investigate the upper critical field anisotropy to higher fields ($H\leq
60$ T), the magnetic field dependence of radio frequency (rf) contactless
penetration depth was measured for applied field both parallel ($\mathbf{H}%
\parallel \mathbf{c}$) and perpendicular ($\mathbf{H}\parallel \mathbf{ab}$)
to the tetragonal \textit{c}-axis. The rf contactless penetration depth
measurements were performed in a 60 T short pulse magnet with a 10 ms rise
and 40 ms decay time. The rf technique has proven to be a sensitive and
accurate method for determining the $H_{c2}$ of superconductors. \cite%
{Mielke2001} This technique is highly sensitive to small changes in the rf
penetration depth ($\sim $ 1-5 nm) in the mixed state. As the magnetic field
is applied, the probe detects the transition to the normal state by tracking
the shift in resonant frequency, which is proportional to the change in
penetration depth as $\Delta \lambda \propto \Delta F/F_{0}$, where $F_{0}$
is 25 MHz in the current setup. Because of the eddy current heating caused
by the pulsed field, small single crystals were chosen, where the sample was
placed in a circular detection coil for $\mathbf{H}\parallel \mathbf{ab}$
and was located on the top surface of one side of the counterwound coil pair
for $\mathbf{H}\parallel \mathbf{c}$. \cite{Coffey2000}$,$\cite%
{Altarawneh2009} For the $\mathbf{H}\parallel \mathbf{c}$ configuration, the
coupling between sample and coil is weaker than that for $\mathbf{H}%
\parallel \mathbf{ab}$, resulting in a smaller frequency shift that is still
sufficient to resolve $H_{c2}(T)$. Details about this technique can be found
in Refs. \cite{Coffey2000}$-$\cite{Altarawneh2008}.

For M\"{o}ssbauer measurements, two cleaved single crystal mosaic samples
were prepared from the same batch of crystals. The first, for
low-temperature work, was prepared by attaching several single crystal
plates to a 12~mm diameter disc of 100 $\mu m$ thick Kapton foil using
Apiezon N grease. Care was taken to ensure that there were no gaps, but
rather minimal overlap between the crystals. This sample was transferred
promptly to a vibration-isolated closed-cycle refrigerator with the sample
held in vacuum. The second sample, for the high-temperature work, was
attached to a $\frac{1}{2}$-inch diameter 10-mil beryllium disc using
diluted GE-7031 varnish before being mounted in a resistively heated oven,
again with the sample in vacuum. While we operated somewhat above the
maximum service temperature of the varnish, the sample was cycled above 250$%
^{\circ }$C three times without any evidence of degradation.

The M\"{o}ssbauer spectra were collected on conventional spectrometers using
50~mCi $^{57}$Co\textbf{Rh} sources mounted on electromechanical drives
operated in constant acceleration mode (on the high-temperature system) and
sine-mode (on the low-temperature system). The spectrometers were calibrated
against $\alpha -$Fe metal at room temperature. The closed-cycle
refrigerator cools to 10~K, with temperature sensing and control using a
calibrated silicon diode mounted on the copper sample stage. Measured
gradients (centre to edge of sample) in the oven are less than 1~K up to
750~K. Control and sensing rely on four, type-K, thermocouples. Temperature
stability in both cases is better than 0.2~K. Spectra were fitted using a
conventional non-linear least-squares minimisation routine to a sum of
equal-width Lorentzian lines. Magnetic patterns were fitted assuming
first-order perturbation in order to combine the effects of the magnetic
hyperfine field ($B_{hf}$) and the electric field gradient.

\section{Crystal growth and stoichiometry}

Although single crystals of K$_{x}$Fe$_{2-y}$Se$_{2}$ could be grown readily
from a melt, various stoichiometries of the single crystals were reported in
literature, with wide ranges of the values of $x,y$ ($0.6\leq x<1$ and $%
0\leq y\leq 0.59$)\cite{Ying}$-$\cite{Fang}$,$\cite{Wangdm}$-$\cite{Zava}.
There is concensus that K$_{x}$Fe$_{2-y}$Se$_{2}$ is of off-stoichiometric
nature and the deficiency of K and Fe strongly influences their electrical
transport properties, tuning the material from insulating to superconducting
state.\cite{Fang}$,$\cite{Wangdm} Different techniques were claimed to be
suscessful in growing single crystals: self-flux growth\cite{Ying},
Bridgeman method.\cite{Wangdm} In order to understand the crystal growth and
obtain well controlled samples, two different ways were tried for growing
single crystals of K$_{x}$Fe$_{2-y}$Se$_{2}$. As-grown crystals were
compared and checked for homogeneity.

The first batch of single crystals of K$_{x}$Fe$_{2-y}$Se$_{2}$ were grown
from K$_{0.8}$Fe$_{2}$Se$_{2}$ melt, as described in Ref. 20. The starting
material was slowly furnace-cooled from 1050 $^{o}$C and dark shiny crystals
could be mechanically separated from the solidified melt, which was
consisted of crystals and fine polycrystalline material. The different
stoichiometry between the starting material and resultant single crystal
clearly implies that this is not simply the cooling of a stoichiometric melt
to form a congruently melting, line compound. A second batch was grown from
a starting composition of KFe$_{3}$Se$_{3}$. The sample was decanted\cite%
{FiskRemi},\cite{PaulFisk} at 850 $^{o}$C after cooled from 1050 $^{o}$C.
This procedure resulted in similar but smaller crystals as the
furnace-cooled samples. It clearly shows that K$_{x}$Fe$_{2-y}$Se$_{2}$
crystals are grown out of a ternary high temperature solution.

The lattice parameters refined from powder x-ray diffraction pattern of the
crystals for both I4/mmm and I4/m space groups were $a=3.8897(8)$\AA\ and $%
c=14.141(3)$\AA . They are in good agreement with the previous reported
values in Ref. 20 ($a=3.8912$\AA , $c=14.139$\AA ). Wave-length dispersive
x-ray spectroscopy (WDS) analysis was performed on both types of crystals to
give a better determination of stoichiometry than the semi-quantitative
Energy Dispersive X-ray (EDX) spectroscopy.\cite{Ying}$-$\cite{Fang}$,$\cite%
{Wangdm}$-$\cite{Zava} The average composition are $%
K:Fe:Se=0.80(2):1.76(2):2.00(3)$ for the furnace-cooled sample and $%
K:Fe:Se=0.79(2):1.85(4):2.00(4)$ for the decanted sample, where the atomic
numbers of K and Fe are normalized to two Se per formula unit and the
standard deviation $\sigma $ is taken as the compositional error and shown
in parentheses after value. We found there is a spread of composition, the
difference between the maximum and minimum values of the measurements, 0.07,
0.06 and 0.10 for K, Fe and Se respectively, for furnace-cooled crystals and
0.04, 0.12 and 0.09 for decanted crystals, roughly within $3\sigma $ of a
normal distribution of random variable. It could be associated with the
broadened superconducting transition, microstructure as seen in scanning
electron microscope and the paramagnetic phase observed in M\"{o}ssbauer
spectroscopy (discussed below). The crystals grown from solution have very
similar composition to the furnace cooled samples, with only a little higher
concentration of Fe, reasonable for a crystal grown out of solution with a
greater excess of Fe-Se.

\section{Physical properties of single crystals of K$_{0.80}$Fe$_{1.76}$Se$%
_{2}$}

\begin{figure}[tbp]
\centerline{\includegraphics[scale=0.45]{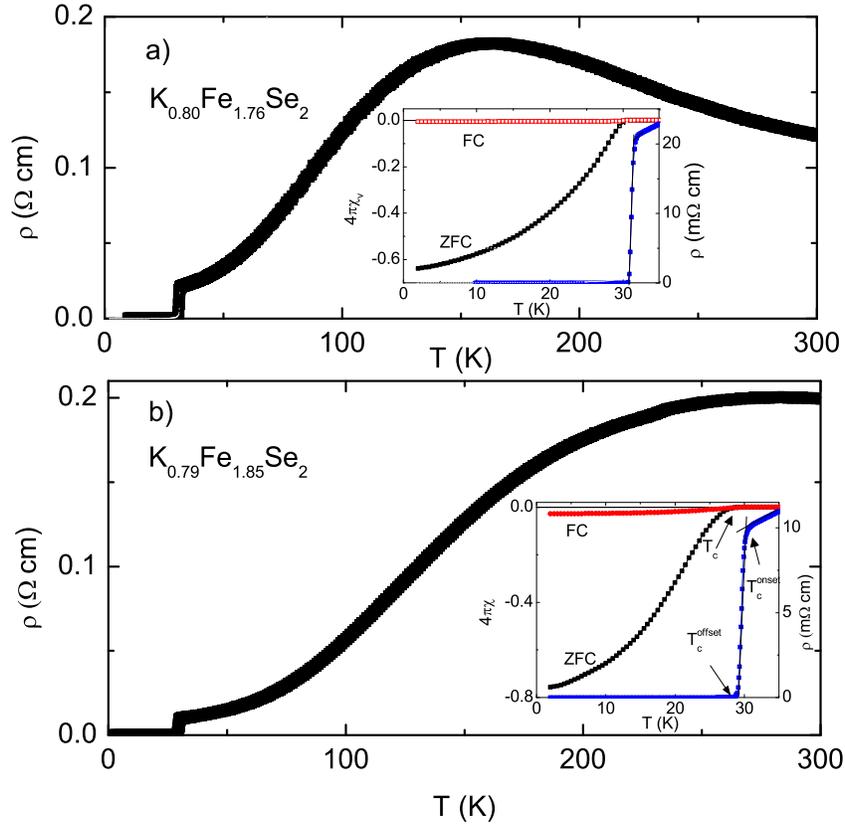}}
\caption{Comparison of the in-plane resistivity, and low temperature
magnetic susceptibility of two types of K$_{x}$Fe$_{2-y}$Se$_{2}$ single
crystals, a) furnace cooled; b) decanted sample. Inset shows the low
temperature region of the resistivity (to the right axis) together with
zero-field-cooled and field-cooled magnetic susceptibility in a field of 50
Oe.}
\label{Fig1.1}
\end{figure}

\begin{figure}[tbp]
\centerline{\includegraphics[scale=0.3]{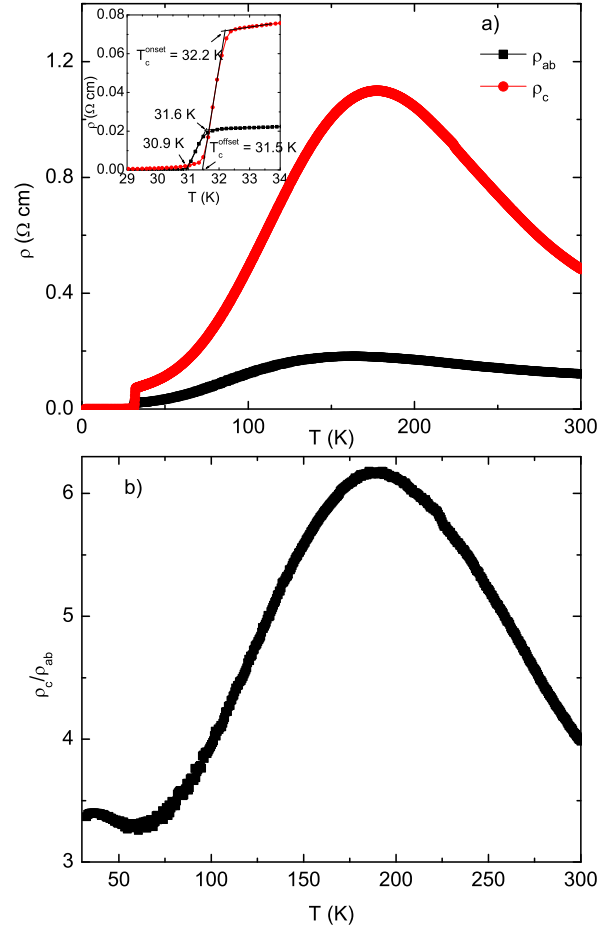}} \vspace*{-0.3cm}
\caption{a) Anisotropic resistivity as a function of temperature. Inset is
an expanded view around the transition. b) Anisotropy of resistivity vs
temperature.}
\label{Fig1.2}
\end{figure}

\begin{figure}[tbp]
\centerline{\includegraphics[scale=0.7]{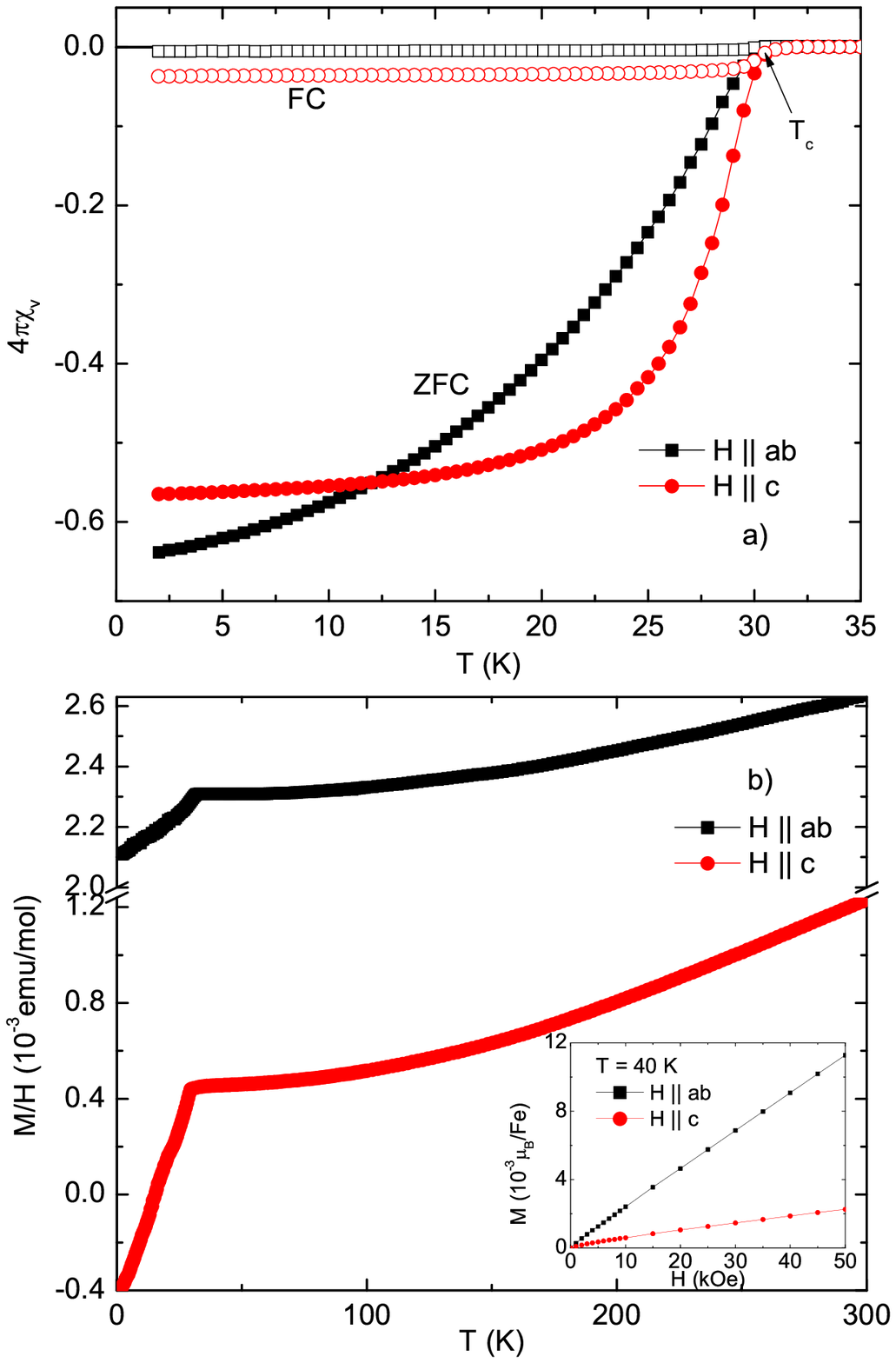}} \vspace*{-0.3cm}
\caption{a) Temperature dependence of low field (H = 50 Oe) magnetic
susceptibility for H$\Vert $\textit{ab} and H$\Vert $\textit{c}; b) Magnetic
susceptibility M/H, measured in 50 kOe for two field directions. Inset shows
field dependence of magnetization at 40 K for both field directions.}
\label{Fig1.3}
\end{figure}

\subsection{Transport and thermodynamic properties}

We compare the temperature dependent electrical resistivity and
magnetization measurements of crystals grown by both the furnace cooled and
decanted methods in Fig. \ref{Fig1.1}. The in-plane resistivity of the
furnace cooled sample is very similar to that of earlier reports.\cite{Ying}$%
,$\cite{Wangdm} Although the superconducting transition temperature inferred
from resistivity are similar ($T_{c}^{offset}=30.9$ K for furnace-cooled
sample and $T_{c}^{offset}=29$ K for decanted sample), the broad resistive
maxima is shifted from 160 K for furnace-cooled sample to 280 K for decanted
sample. Wang \textit{et al.} showed that the position of the hump is
sensitive to Fe deficiency.\cite{Wangdm} With decreasing Fe deficiency, the
hump shifts to higher temperature. This observation agrees well with the WDS
result, which shows smaller Fe deficiency in the decanted samples. Given the
small difference of both types of single crystals and the similarity to
samples from earlier reports of the furnace-cooled samples, for the rest of
this paper we will focus their fuller characterization.

Anisotropic resistivity as a function of temperature is shown in Fig. \ref%
{Fig1.2}a. It is clear that there is a broad maximum peak around 160 K for $%
\rho _{ab}$ and 180 K for $\rho _{c}$. The difference of maximum positions
suggest that they result from a crossover rather than transition. The
anisotropy is probably due to the layered structure of K$_{0.80}$Fe$_{1.76}$%
Se$_{2}$. Figure \ref{Fig1.2}b shows the anisotropy $\rho _{c}/\rho _{ab}$,
reaches the maximum of 6 around 180 K and decreases to 4 around 300 K. It is
comparable to the anisotropy of AFe$_{2}$As$_{2}$.\cite{Tanatar} But a much
larger resistivity anisotropy of 30-45 was reported in (Tl,K)Fe$_{x}$Se$_{2}$%
\cite{Wanghd}, this may imply that the specific composition influences
carrier tunneling significantly. An expanded view around the superconducting
transition is shown in the inset to Fig. \ref{Fig1.2}a. For both of the
current directions, the transition width is about 0.7 K, but the inferred $%
T_{c}$ value from $\rho _{c}$ is slightly higher than that of $\rho _{ab}$. 
\begin{figure}[tbp]
\centerline{\includegraphics[scale=0.4]{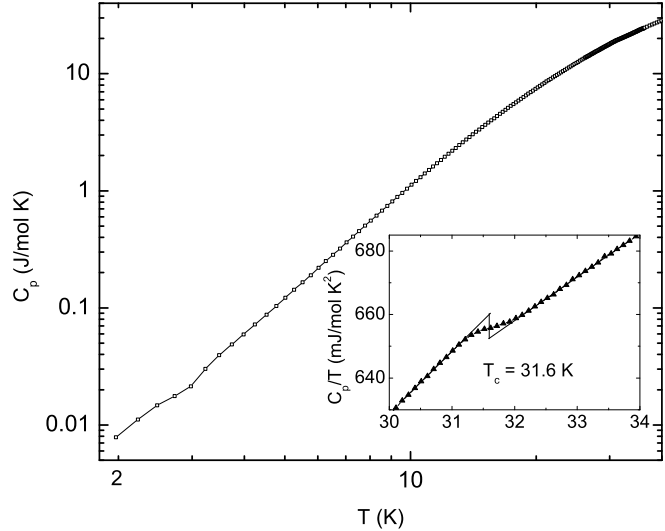}} \vspace*{-0.3cm}
\caption{Specific heat as a function of temperature on a log-log plot. Inset
shows the heat capacity jump at the superconducting transition. The solid
line is an isoentropic estimate of $T_{c}$ and $\Delta C_{p}$.}
\label{Fig1.4}
\end{figure}

\begin{figure}[tbp]
\centerline{\includegraphics[scale=0.4]{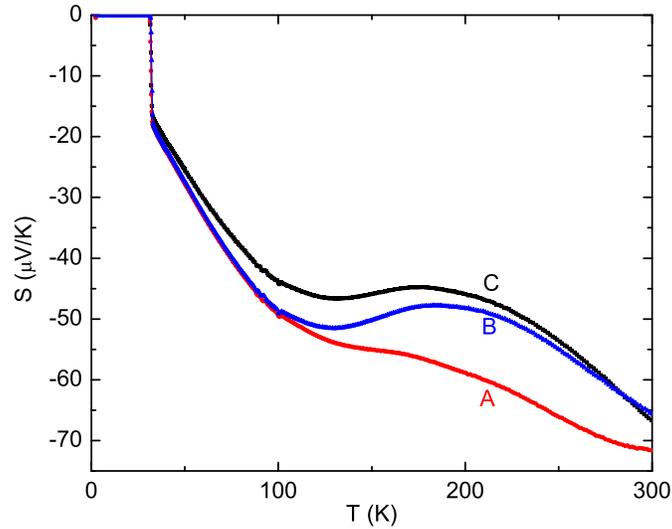}} \vspace*{-0.3cm}
\caption{Thermoelectric power as a function of temperature. Samples A and B
use silver paste as contact (contact resistance $\sim 1-3$ k$\Omega $).
Sample C uses silver wires attached by In-Sn solder as contact (contact
resistance $\sim 200$ $\Omega $).}
\label{Fig1.5}
\end{figure}

\begin{figure}[tbp]
\centerline{\includegraphics[scale=1.1]{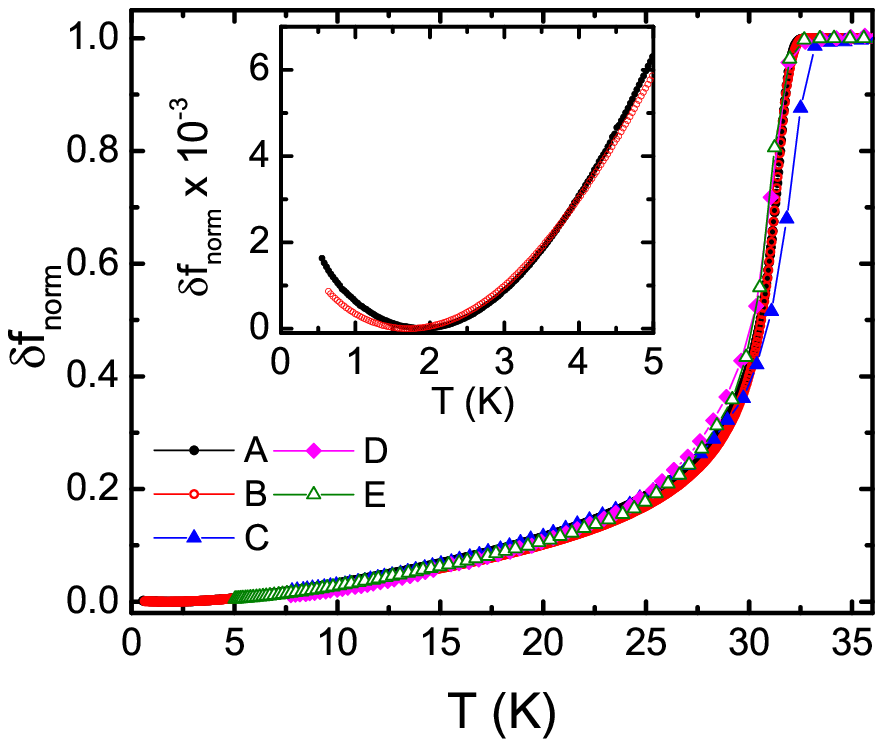}} \vspace*{-0.3cm}
\caption{Normalized London penetration depth expressed via resonant
frequency shift, $\Delta f_{norm}=(f(T)-f(T_{c}))/(f(T_{c})-f(T_{min}))$
proportional to magnetic susceptibility. $f(T_{min})$ is the resonant
frequency at the lowest temperature $\simeq $ 0.5 K. $f(T_{c})$ is the
frequency in the normal state right above $T_{c}$. Inset shows an upturn,
presumably due to paramagnetic ions and/or impurities below 2 K from two
samples A and B.}
\label{Fig1.6}
\end{figure}

Figure \ref{Fig1.3}a shows the magnetic susceptibility of K$_{0.80}$Fe$%
_{1.76}$Se$_{2}$ for two directions of an applied field of 50 Oe. For $%
\mathbf{H}$ $\Vert $ $\mathbf{ab}$\textit{,} the zero-field-cooled (ZFC)
curve decreases slowly with temperature and for $\mathbf{H}$ $\Vert $\textbf{%
\ }$\mathbf{c}$ the transition becomes sharper. Similar behavior can be seen
in Tl$_{0.58}$Rb$_{0.42}$Fe$_{1.72}$Se$_{2}$.\cite{Wanghd} This temperature
dependence of the ZFC curve is similar to an inhomogeneous superconductor
with a range of transition temperatures and may be related to the small
spread of stoichiometry found in WDS data. Both of the ZFC curves in Fig. %
\ref{Fig1.3}a approach -0.6 consistent with substantial shielding and $T_{c}$
inferred from both curves is the same, $T_{c}=30.1\pm 0.1$ K, within
experimental error. The magnetic susceptibility M/H (H = 50 kOe) as a
function of temperature for both field directions is shown in Fig. \ref%
{Fig1.3}b. Similar temperature dependence is observed for both field
directions, i.e. M/H decreases almost linearly with decreasing temperature
above 150 K and shows a sudden drop below 30 K associated with
superconductivity. $\chi _{ab}$ is clearly larger than $\chi _{c}$ over the
whole temperature range. No anomalies in magnetic susceptibility can be
correlated with the broad maxima in resistivity. The linear field dependence
of magnetization at 40 K for both directions (Fig. \ref{Fig1.3}b inset)
indicates that there are no ferromagnetic impurities, and the
non-Curie-Weiss like temperature of the susceptibility indicates that the
system might be deep in an antiferromagnetic state, consistent with what was
suggested for Cs$_{0.8}$Fe$_{2}$Se$_{1.96}$\cite{Sher} and K$_{0.8}$Fe$%
_{1.6} $Se$_{2}$\cite{Weibao}.

Specific heat data was collected to verify the bulk thermodynamic nature of
the superconducting transition. $C_{p}$ vs $T$ at low temperature is shown
in Fig. \ref{Fig1.4} on a log-log plot. In the superconducting state, below
15 K, $C_{p}$ roughly follows a $T^{3}$ power law. This implies a dominant
phonon contribution and a very small electronic term. $C_{p}/T$ vs \textit{T}
is plotted in the inset for $T\sim T_{c}$ and a clear jump of specific heat
associated with the superconducting transition at 31.6 K is seen and $\Delta
C_{p}/T_{c}=7.7$ mJ/mol K$^{2}$, can be identified. The jump is
substantially less than jump seen for K-doped Ba122 samples; in comparison
to the $\Delta C_{p}/T_{c}$ versus $T_{c}$ presented by Bud'ko \textit{et al.%
}\cite{BNC}, this jump is $\sim $ 15\% of what would be expected from a
doped 122 material.\cite{NiBaK}

The thermoelectric power (TEP) as a function of temperature is shown in Fig. %
\ref{Fig1.5}. Three different samples with different eletrical and thermal
contact were shown to have consistent $T_{c}=31.6$ K inferred from $S(T)=0$.
The data for three samples are similar over the whole temperature range. The
origins of local minimum and maximum found between $100-200$ K are not
clear, but it is very likely that they are associated with the multiband
structure of K$_{0.80}$Fe$_{1.76}$Se$_{2}$ and the crossover (metal-like at
low temperature) observed in resistivity. The negative sign of the
thermopower indicates that electron like carriers are dominant, thus in
agreement with the observation of electron only pockets at the Fermi surface
by Angle Resolved Photoemission Spectroscopy (ARPES).\cite{Qian} The large
absolute value of S above 50 K is similar to TEP data observed for Co-doped
BaFe$_{2}$As$_{2}$\cite{EDMun} and has been reproduced by other, recent TEP
measurements. \cite{KFWang}$,$\cite{YJYan} 
\begin{figure}[tbp]
\centerline{\includegraphics[scale=1]{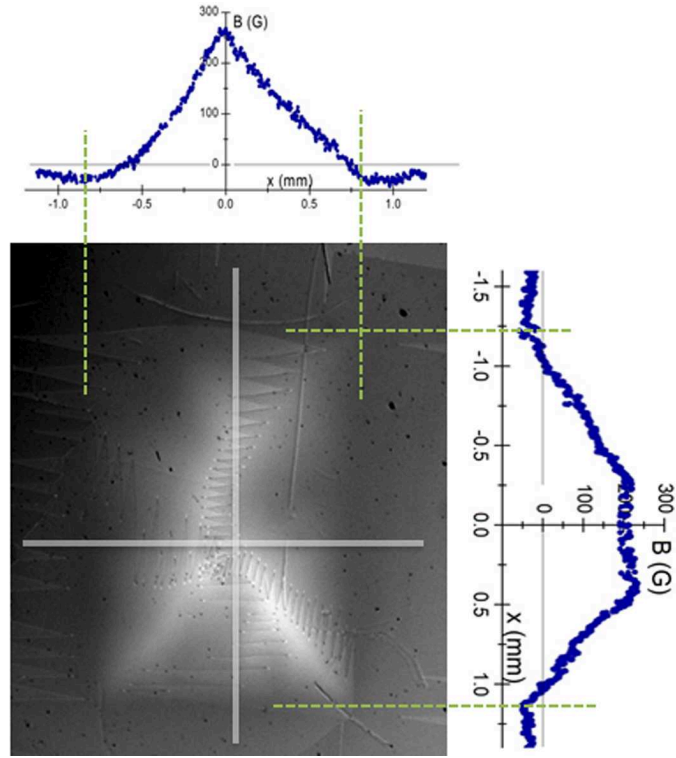}} \vspace*{-0.3cm}
\caption{Magneto-optical image of single crystal K$_{0.80}$Fe$_{1.76}$Se$%
_{2} $. Grey lines show the cuts along which the magnetic induction was
measured and shown on the side panels.}
\label{Fig1.7}
\end{figure}

\subsection{\protect\bigskip London penetration depth and magneto-optical
imaging}

London penetration depth measurements with good reproducibility were
performed on several single crystal samples. Figure \ref{Fig1.6} shows the
normalized frequency shift, proportional to differential magnetic
susceptibility, $\delta f_{norm}=(f(T)-f(T_{c}))/(f(T_{c})-f(T_{min}))$,
where $f(T_{min})$ is the resonant frequency at the lowest temperature $%
\simeq $ 0.5 K and $f(T_{c})$ is the frequency in the normal state right
above $T_{c}$. Consistent measurements on several samples indicate little or
no variation within the batch. The transition itself is quite unusual - it
shows quite a sharp onset, but then is smeared almost over the entire
temperature interval. This also might be due to the small variation of the
stoichiometry or impurities. It is also possible that the observed behavior
is indicative of strongly anisotropic gap function or even nodes. In
addition, there is a clear upturn at low temperatures. It has been shown in
both, high-Tc cuprates\cite{Ruslan0} and 1111 pnictides\cite{Martin} that
this upturn can be caused by the paramagnetic ions.

Magneto-optical imaging can shed more light on the homogeneity of the
superconducting state (at least for length scales larger than the wavelength
of optical light) and gives a rough estimate critical current density. A
magneto-optical image of a trapped flux in a field-cooled sample is shown in
Fig. \ref{Fig1.7}. We did not observe any noticeable Meissner expulsion,
similar to other 122 pnictides.\cite{Ruslan1} When magnetic field was turned
off, it revealed a typical \textquotedblleft Bean\textquotedblright\ roof,
again similar to other pnictide superconductors.\cite{Ruslan2}$,$\cite%
{Ruslan3} As can be seen in Fig. \ref{Fig1.7}, the magnetic flux
distribution is relatively uniform; however, some macroscopic variations
(upper left corner) might indicate some smooth variation of stoichiometry
across the sample and may help to explain the broadened transition curves.
In order to quantify the critical state, Fig. \ref{Fig1.7} also shows
profiles of the magnetic induction taken along two lines (shown in the
figure). The remanence reaches about 250 Oe. A simple one-dimensional
estimate, using%
\begin{equation*}
\frac{4\pi }{c}j_{c}=\frac{dB}{dx}
\end{equation*}

gives:

\begin{equation*}
j_{c}=\frac{250}{0.77}\frac{10}{4\pi }\approx 2.6\times 10^{3}\text{ A/cm}%
^{2}
\end{equation*}

This shows that the current samples cannot support large critical current
density even at low temperatures. Similar numbers are estimated from the
magnetization measurements.\cite{Cedomir}

\subsection{Anisotropic H$_{c2}$(T)}

\begin{figure}[tbp]
\centering
\includegraphics[width=1.0\linewidth]{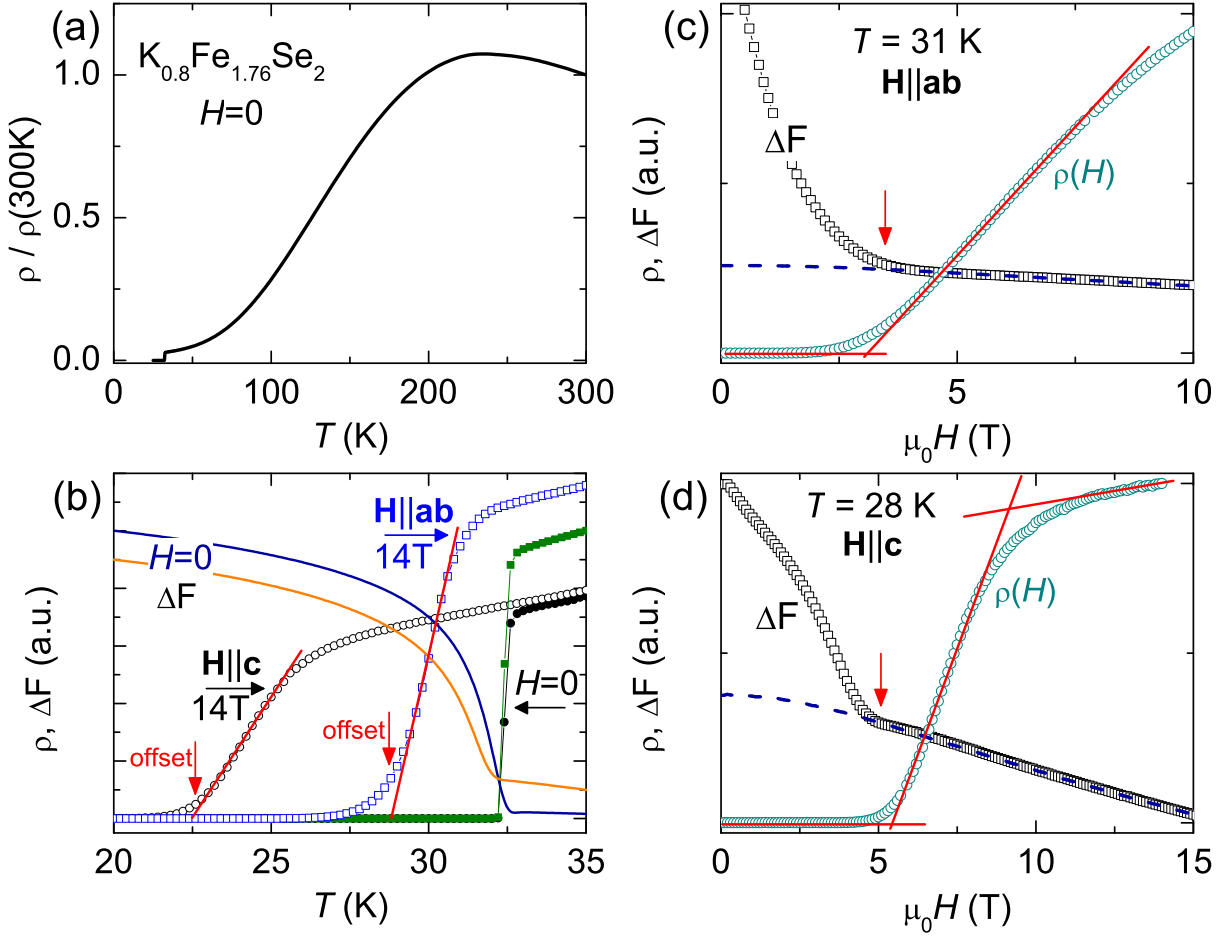}
\caption{(a) Temperature dependence of the normalized $\mathbf{ab}$-plane
resistivity $\protect\rho (T)$ of the K$_{0.8}$Fe$_{1.76}$Se$_{2}$ single
crystal at $H$ = 0, where $\protect\rho $(300K) = 0.12 $\Omega $ cm. (b) Low
temperature region of the resistance for two samples at $H$ = 0 (closed
symbols) and 14 T (open symbols) and the warming curves of rf shift ($\Delta 
$F) for two samples (solid lines). Vertical arrows indicate $T_{c}^{\mathtt{%
offset}}$ and lines on the top of 14 T data are guide to the eye. (c)
Comparison of the $\mathbf{ab}$-plane resistance $R(H)$ and $\Delta $F for $%
\mathbf{H}\parallel \mathbf{ab}$ at $T$ = 31 K. (d) Comparison of the $%
\mathbf{ab}$-plane resistance $R(H)$ and $\Delta $F for $\mathbf{H}\parallel 
\mathbf{c}$ at $T$ = 28 K. The dashed lines in (c) and (d) are the $\Delta $%
F taken at $T$ = 35 K as a normal state, background signal. The solid lines
in (c) and (d) are guides to the eye for offset and onset criteria of $H_{c}$
and vertical arrows indicate the deviation of $\Delta F$ from the background
signal (see text).}
\label{Fig1.8}
\end{figure}

The anisotropic $H_{c2}$ curves for K$_{0.8}$Fe$_{1.76}$Se$_{2}$ are
inferred from measurements of magnetoresistance (for $\mathbf{H}$ $\leq $ 14
T) and from high magnetic field measurements of radio frequency (rf)
contactless penetration depth for applied field both parallel ($\mathbf{H}%
\parallel \mathbf{c}$) and perpendicular ($\mathbf{H}\parallel \mathbf{ab}$)
to the tetragonal \textit{c}-axis. Figure \ref{Fig1.8}a shows the
temperature dependence of the normalized resistivity for the K$_{0.8}$Fe$%
_{1.76}$Se$_{2}$ sample. The offset and zero-resistance ($R<3\times
10^{-5}\Omega $) temperatures were estimated to be $T_{c}^{offset}$ $\simeq $
32.2 K and $T_{c}^{\mathtt{zero}}$ $\simeq $ 32 K, respectively, as shown in
Fig. \ref{Fig1.8}b. The solid lines in Fig. \ref{Fig1.8}b are warming curves
of the rf shift ($\Delta $F) at $H$ = 0 for two different samples. As the
temperature decreases, the rf shift suddenly increases at $T_{c}$, where $%
T_{c}$ = 32 and 32.4 K for two samples were determined from d$\Delta $F/d$T$%
. A clear anisotropy in the response of the superconductivity under applied
fields was observed between $\mathbf{H}\parallel \mathbf{ab}$ and $\mathbf{H}%
\parallel \mathbf{c}$ as shown in Fig. \ref{Fig1.8}b for $H$ = 14 T curves.
To compare the superconducting transition between resistance and $\Delta $F
measurement, resistance data measured in a superconducting magnet and $%
\Delta $F taken in pulsed magnetic fields at $T$ = 31 K for $\mathbf{H}%
\parallel \mathbf{ab}$ and at $T$ = 28 K for $\mathbf{H}\parallel \mathbf{c}$
are plotted in Figs. \ref{Fig1.8}c and d, respectively. As shown in the
figures, the deviation from the background signal of $\Delta $F is close to
the $H_{c}^{\mathtt{offset}}$ criterion of the resistance curves.

\begin{figure}[tbp]
\centering
\includegraphics[width=1.0\linewidth]{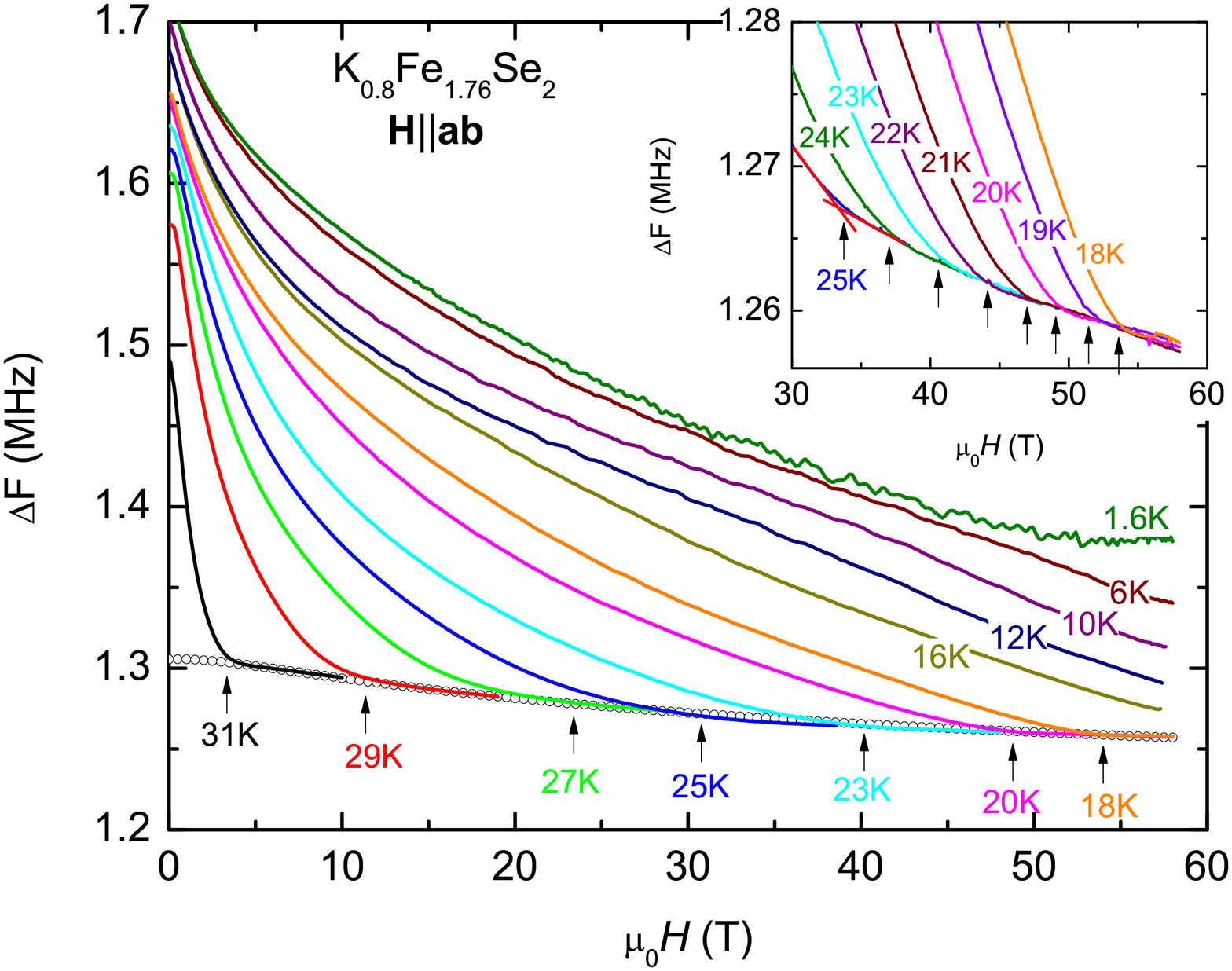}
\caption{Frequency shift ($\Delta$F) as a function of magnetic field for $%
\mathbf{H}\parallel\mathbf{ab}$ at selected temperatures. Open symbols are $%
\Delta$F taken at $T$ = 35 K as a normal state, background signal. The
arrows indicate $H_{c2}$ determined from the point deviating from background
signal. Inset shows the low temperature data close to $H_{c}$. The straight
lines on the $T$ = 25 K curve are guides to the eye for determining the
point at which the rf signal intercepts the slope of the normal state
background.}
\label{Fig1.9}
\end{figure}

\begin{figure}[tbp]
\centering
\includegraphics[width=1.0\linewidth]{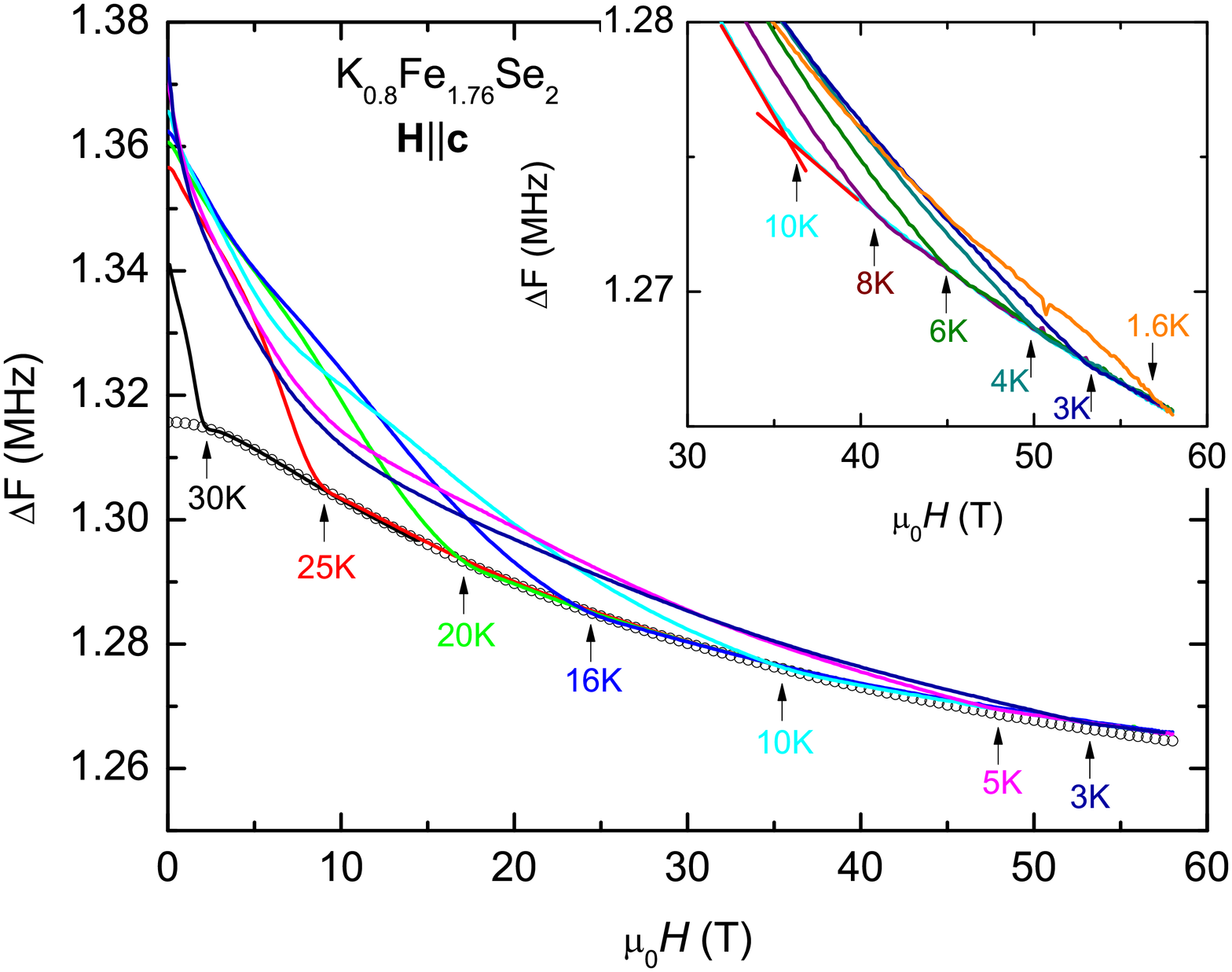}
\caption{Frequency shift ($\Delta$F) as a function of magnetic field for $%
\mathbf{H}\parallel\mathbf{c}$ at selected temperatures. Open symbols are $%
\Delta$F taken at $T$ = 35 K as a normal state, background signal. The
arrows indicate $H_{c2}$ determined from the point deviating from background
signal. Inset shows the low temperature data close to $H_{c}$. The straight
lines on $T$ = 10 K curve are guides to the eye for determining the point at
which the rf signal intercepts the slope of the normal state background.}
\label{Fig1.10}
\end{figure}

The $\Delta $F vs $H$ plots shown in Figs. \ref{Fig1.9} and \ref{Fig1.10}
can be used to infer the temperature dependence of the upper critical field $%
H_{c2}(T)$ by simply taking the first point deviating from the normal state
background. Arrows in Figs. \ref{Fig1.9} and \ref{Fig1.10} indicate the
determined $H_{c2}$. The difference between the $H_{c2}$ values determined
by the first deviation and slope change point criteria was used to determine
the $H_{c2}$ error bar size. 

\begin{figure}[tbp]
\centering
\includegraphics[width=1.0\linewidth]{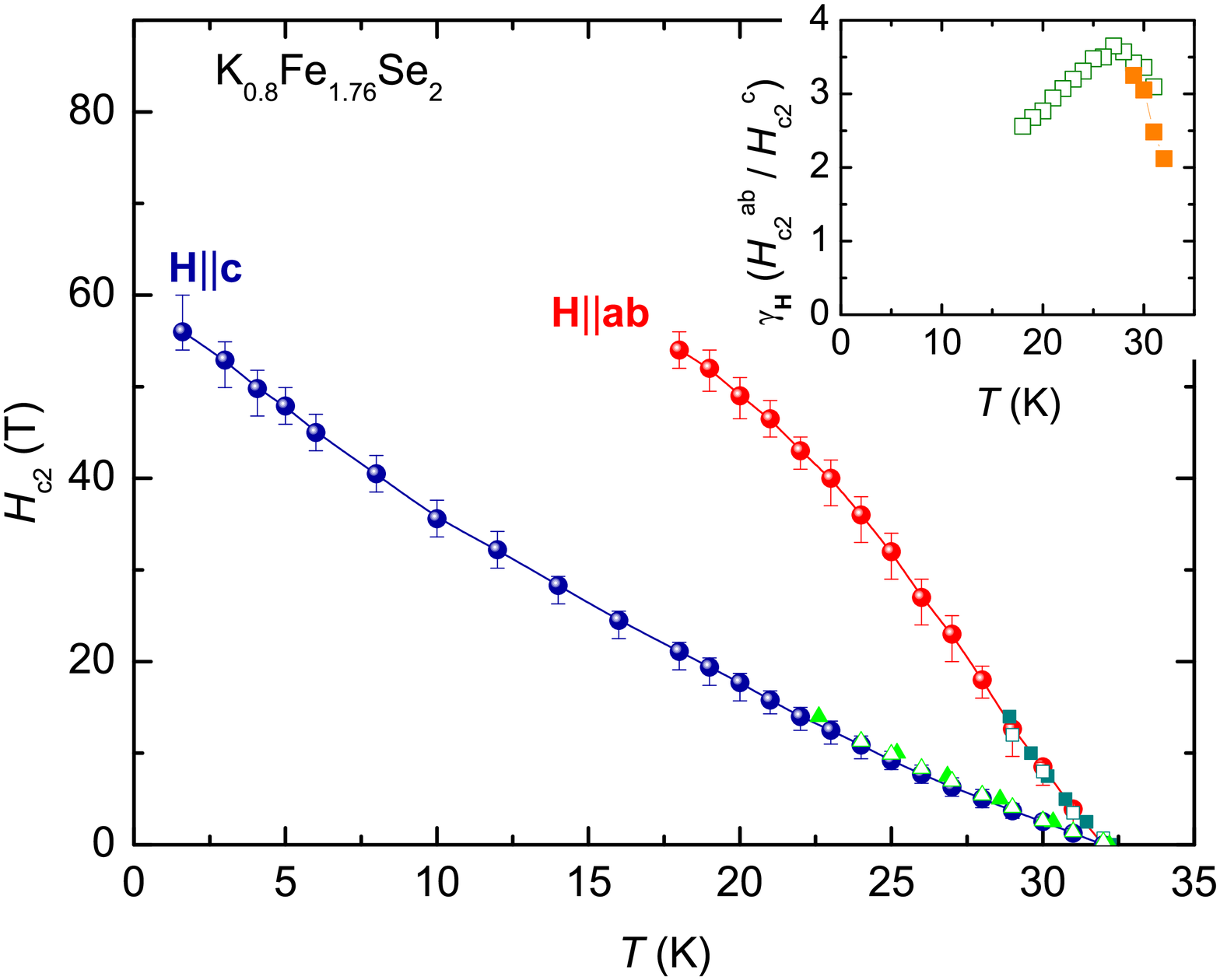}
\caption{Anisotropic $H_{c2}$($T$) for K$_{0.8}$Fe$_{1.76}$Se$_{2}$ single
crystals. Solid circles are obtained from the pulsed field rf shift
measurements and closed (open) square and triangle symbols are taken from
temperature (magnetic field)-dependent resistance measurements. Inset shows
the temperature dependence of the anisotropy $\protect\gamma _{H}$ = $%
H_{c2}^{ab}/H_{c2}^{c}$ as determined from pulsed field rf shift (open
squares) and resistance (solid squares) measurements.}
\label{Fig1.11}
\end{figure}

The $H_{c2}$($T$) curves for both $\mathbf{H}\parallel \mathbf{ab}$ ($%
H_{c2}^{ab}$) and $\mathbf{H}\parallel \mathbf{c}$ ($H_{c2}^{c}$) in K$%
_{0.8} $Fe$_{1.76}$Se$_{2}$ are plotted in Fig. \ref{Fig1.11}, as determined
from the $H\leq 14$ T resistance and from the $H\leq 60$ T data taken from
the down sweep of pulsed field magnetic field rf measurements. The curvature
of $H_{c2}$($T$) has been reported to vary depending on the criteria used to
determine $H_{c2}$, for example in the case of highly two dimensional, high-$%
T_{c}$ cuprate superconductors. \cite{Ando1999} In this study, the shape of $%
H_{c2}$ curves does not change qualitatively when $H_{c2}$ is defined by
different criteria or even different measurements. On the other hand, the
shapes of the upper critical field curves for $\mathbf{H}\parallel \mathbf{ab%
}$ and $\mathbf{H}\parallel \mathbf{c}$ clearly do not manifest the same
temperature dependence. As is evidenced from Fig. \ref{Fig1.11}, a
conventional linear field dependence of $H_{c2}$ is observed close to the $%
T_{c}$, with clearly different slopes for the two field orientations. In the
low field region the $H_{c2}$ curves are consistent with earlier studies.%
\cite{Wangdm}$,$\cite{Mizuguchi} Towards higher fields, $H_{c2}^{c}$($T$)
presents an almost linear temperature dependence down to 1.5 K, whereas the
curve of $H_{c2}^{ab}$($T$) has a tendency to saturate. The anisotropy
parameter, $\gamma _{H}\equiv $ $H_{c2}^{ab}/H_{c2}^{c}$, is about $\sim $ 2
near $T_{c}$, but shows a maximum around 27 K with $\gamma _{H}\sim $ 3.6,
and decreases considerably for lower temperatures. In all known examples so
far, the temperature dependence of $\gamma _{H}$ was opposite to that of $%
\gamma _{\lambda }\equiv \lambda _{c}/\lambda _{ab}$. It would be
interesting to examine $\gamma _{\lambda }(T)$ in this material, in
particular to see if it goes through a minimum at $\sim 27$ K.

The zero temperature limit of $H_{c2}$ can be estimated by using the
Werthamer-Helfand-Hohenberg (WHH) theory\cite{Werthamer1966}, which gives $%
H_{c2}$ = 0.69$T_{c}$(d$H_{c2}$/d$T$)$\mid _{T_{c}}$. The value of $H_{c2}$%
(0) for $\mathbf{H}\parallel \mathbf{ab}$ and $\mathbf{H}\parallel \mathbf{c}
$ were estimated to be $\sim $ 102 T and $\sim $ 31 T respectively, where $%
T_{c}$ = 32 K, d$H_{c2}^{ab}$/d$T$ $\sim $ -4.6 T/K and d$H_{c2}^{c}$/d$T$ $%
\sim $ -1.4 T/K were used. Clearly these values do not capture the salient
physics for this compound. On the other hand, in the simplest approximation,
the Pauli limit ($H_{P}$) is given by 1.84$T_{c}$, \cite{Clogston1962}$-$%
\cite{Maki1966} giving $H_{P}$ $\sim $ 59 T. This low temperature value of $%
H_{c2}$ may indeed capture some of the basic physics associated with K$%
_{0.8} $Fe$_{1.76}$Se$_{2}$. To explain the observed $H_{c2}$ curves in
detail, a more complete theoretical treatment is needed, one that does not
exclude the strong electron-phonon coupling and multiband nature of Fe-based
compounds. Anisotropic superconducting coherence length can be calculated
using $H_{c2}^{ab}=\frac{\phi _{0}}{2\pi \xi _{ab}\xi _{c}}$ and $H_{c2}^{c}=%
\frac{\phi _{0}}{2\pi \xi _{ab}^{2}}$. \cite{Poole2000} If $H_{c2}^{c}=60$ T
and $H_{c2}^{ab}$ is assumed to be between 60 and 100 T, then $\xi _{ab}\sim
2.3$ nm, and 1.4 nm $\lesssim \xi _{c}\lesssim 2.3$ nm.

The behavior of $H_{c2}$($T$) for K$_{0.8}$Fe$_{1.76}$Se$_{2}$ is found to
be very similar to that of several 122 systems as well as doped FeSe.\cite%
{Altarawneh2008}$,$\cite{Ni2008}$-$\cite{Kano2009} It should be noted that
the $H_{c2}$ curves for two orientations in K-doped BaFe$_{2}$As$_{2}$
system seem to cross at low temperature due to the flattening of $%
H_{c2}^{ab} $($T$) curve, \cite{Altarawneh2008}$,$\cite{Yuan2009}
additionally, the $H_{c2}$ curves for FeTe$_{0.6}$Se$_{0.4}$ shows a
crossing between $\mathbf{H}\parallel \mathbf{ab}$ and $\mathbf{H}\parallel 
\mathbf{c}$ curves below 4.5 K because of the subsequent flattening of the $%
H_{c2}^{ab}$ curve at low temperatures.\cite{Khim2010}$,$\cite{Fang2010R}
However in the Co-doped system, the anisotropic $H_{c2}(T)$ curves do not
show such crossing, \cite{Ni2008}$,$\cite{Kano2009} a result similar to what
was found in this study. Thus, an intriguing feature of $H_{c2}$($T$) curves
for Co- and K-doped BaFe$_{2}$As$_{2}$, FeTe$_{0.6}$Se$_{0.4}$ and K$_{0.8}$%
Fe$_{1.76}$Se$_{2}$ systems is that the anisotropy near $T_{c}$ is as large
as 3 but drops towards $\sim $ 1 as $T\rightarrow 0$ K. The $H_{c2}(T)$
anisotropy in K$_{0.8}$Fe$_{1.76}$Se$_{2}$ is particularly noteworthy given
that it exists deep within an antiferromagnetically ordered state. \cite%
{Weibao2}$,$\cite{Ryan2011} In the case of Co-doped Ba122, $\gamma
_{H}(T)\sim 1$ when $T_{c}<T_{N}$ with clear anisotropy only emerging when
the antiferromagnetic state is suppressed. \cite{Ni2008}

\subsection{$^{57}$Fe M\"{o}ssbauer spectroscopy}

\begin{figure}[h]
\centerline{\includegraphics [height=12cm]{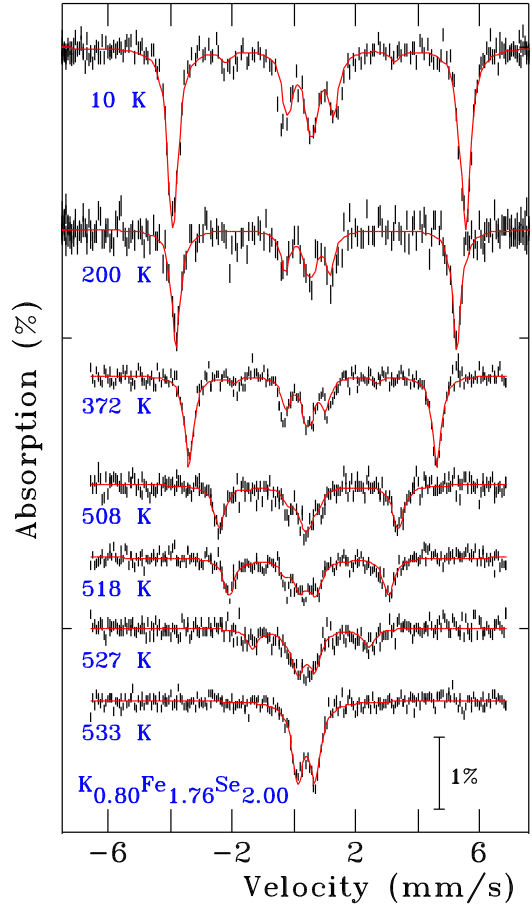}}
\caption{(color online) $^{57}$Fe M\"{o}ssbauer spectra of $\mathrm{%
K_{0.80}Fe_{1.76}Se_{2.00}}$ showing the evolution of the magnetic ordering
on heating from 10~K (well below T$_{c}\sim 30$~K) to 533~K where the
material is paramagnetic. The extreme weakness of the $\Delta m_{I}=0$
transitions in the ordered state indicates that the moments are almost
parallel to the crystal c-axis (see text), while the growth of a central
paramagnetic component above 500~K is characteristic of a first order
magnetic transition. Solid lines are fits as described in the text.}
\label{Fig1.12}
\end{figure}

Room temperature neutron diffraction studies of Cs$_{y}$Fe$_{2-x}$Se$_{2}$ 
\cite{Pomj1919} and A$_{y}$Fe$_{2-x}$Se$_{2}$ (A = Rb, K) \cite{Pomj3380}
have suggested that the iron moments may be much smaller ($\sim $2.5~$\mu
_{B}$/Fe) and also that the magnetic structure may be far more complex than
initially suggested, with the iron atoms being distributed among two
(magnetically) inequivalent sublattices and carrying very different magnetic
moments. Moreover, even the ordering \emph{direction} has been questioned
and it is possible that the iron moments may lie in the \textit{ab}-plane,
at least for Cs$_{y}$Fe$_{2-x}$Se$_{2}$ \cite{Pomj1919}, rather than
parallel to the \textit{c}-axis as initially suggested \cite{Weibao}. Given
the these questions surrounding the magnetic ordering of the iron moments in
the A$_{y}$Fe$_{2-x}$Se$_{2}$ system, we have undertaken a $^{57}$Fe M\"{o}%
ssbauer study K$_{0.8}$Fe$_{1.76}$Se$_{2}$. Whereas M\"{o}ssbauer
spectroscopy cannot be used to determine magnetic structures directly, it is
a quantitative local probe that can be used to set hard limits on possible
structures. As we will show below, the observation of a single, well-split
magnetic component allows us to rule out any structure in which the iron
sub-lattice is further subdivided into multiple, inequivalent sites, and the
scale of the splitting ($\sim $29~T at 10~K) is consistent with the 3.31~$%
\mu _{B}$ moment derived from neutron scattering \cite{Weibao}

Several conclusions can be reached simply by inspection of the spectrum
taken at 10~K (Fig. \ref{Fig1.12}). The spectrum is dominated by a single,
well-split, magnetic component. This confirms that K$_{0.8}$Fe$_{1.76}$Se$%
_{2}$ is indeed magnetically ordered in the superconducting state (recall $%
T_{c}\sim 30$~K for this sample). A small quadrupole splitting of 0.33$\pm $%
0.02~mm/s is present and the linewidth (full width at half maximum) is 0.200$%
\pm $0.007~mm/s, slightly larger than our typical instrumental width of
0.15~mm/s. The single magnetic component allows us to rule out any magnetic
structures involving multiple iron sub-sites with moments that differ by
more than a few percent. As we will show below, the large hyperfine field ($%
B_{hf}\sim $29~T) is inconsistent with a small iron moment and so places
further limits on possible magnetic structures. Finally, two of the lines
normally present in a magnetically split $^{57}$Fe M\"{o}ssbauer spectra,
are essentially absent from the 10~K pattern.

A magnetic field at the $^{57}$Fe nucleus, either externally applied or
transferred from an ordered moment on the iron atom, lifts the degeneracy of
the nuclear states and, in combination with the selection rules for the $%
\frac{3}{2}\rightarrow \frac{1}{2}$ transition, leads to a six-line pattern
with intensities of 3:R:1:1:R:3 (counting from left to right in Fig. \ref%
{Fig1.12}). For a powder sample, R=2, however if there is a unique angle, $%
\theta $, between the magnetic field and the direction of the $\gamma -$beam
used to record the spectrum, then the intensity, R, of the $\Delta m_{I}=0$
transitions is given by: 
\begin{equation*}
R=\frac{4\sin ^{2}\theta }{1+\cos ^{2}\theta }
\end{equation*}%
R = 0 implies that $\theta $ is also zero so that the magnetic field, and by
extension, the moments that lead to it, are parallel to the $\gamma -$beam.
Since the sample consists of an ab-plane mosaic of single crystals, setting $%
\theta =0$ means that the magnetic ordering direction is parallel to the 
\textit{c}-axis, ruling out any magnetic structures that involve planar
ordering of the iron moments. We note that R is a relatively soft function
of $\theta $ near zero, and a free fit to the intensity of the $\Delta
m_{I}=0$ transitions is consistent with an angle of 18$\pm 4^{\circ }$, and
leads to a slight improvement in the least square fit error, $\chi ^{2}$,
for the fit. Such an angle would not be consistent with a purely planar
ordering of the iron moments (indeed, if the ordering were planar, then R
would be 4, and the $\Delta m_{I}=0$ transitions would provide the strongest
features in the spectrum) but it is too large to be dismissed as being due
to a simple mis-alignment of the mosaic. This suggests that there is a small
canting of the antiferromagnetic structure away from the \textit{c}-axis.

Estimating the iron moment from the observed hyperfine field requires some
care as the scaling is imperfect at best\cite{dubiel}. However, some data
exist on binary iron--chalcogenides that can be used as a guide (Table~I).
If we use the factor of 6.2~T/$\mu _{B}$ for Fe$_{7}$Se$_{8}$ with our
measured $B_{hf}$ of 29.4~T we obtain a rather large estimate of 4.7~$\mu
_{B}$/Fe for the iron moment in this system. This is significantly larger
than the 3.31~$\mu _{B}$/Fe reported on the basis of neutron diffraction\cite%
{Weibao}, however it does suggest that the iron moment is indeed substantial
as even the larger conversion factor for the sulphide yields 3.5~$\mu _{B}$%
/Fe. If we assume that $B_{hf}$ is at least proportional to the iron moment,
then we can use the observed change in $B_{hf}$ between 10~K and 293~K to
scale the 3.31~$\mu _{B}$/Fe observed at 11~K\cite{Weibao} to get an
estimate of 3.0~$\mu _{B}$/Fe for the moment at room temperature for
comparison with the much smaller value of 2.55~$\mu _{B}$/Fe reported by
Pomjakushin \textit{et al..}\cite{Pomj3380} However, the strong temperature
dependence of magnetic signal noted by Bao \textit{et al.}\cite{Weibao}
suggests a very rapid decline in ordered moment to about 2.8~$\mu _{B}$/Fe
by room temperature. It is possible that much of the variation may be
intrinsic to the material and its variable stoichiometry, so that combined
measurements on a well characterised sample will be needed to settle this.

\begin{table}[h]
\caption{Average hyperfine fields (B$_{hf}$) derived from $^{57}$Fe M\"{o}%
ssbauer spectroscopy and average iron moments derived from neutron
diffraction for approximately equi-atomic iron--chalcogenide compounds with
estimated field--moment conversion factors. The Fe--Te system exhibits
significant variability and measurements have yet to be made on common
samples making the conversion factor unreliable. There is however a clear
trend to lower values in the sequence S$\rightarrow $Se$\rightarrow $Te. }
\label{tab:conv}%
\begin{tabular}{lccc}
\hline\hline
Compound & Average & Average & Conversion \\ 
& B$_{hf}$ & moment & Factor \\ 
& (T) & $\mu _{B}$/Fe & T/$\mu _{B}$ \\ \hline
&  &  &  \\ 
Sulphides &  &  &  \\ 
Fe$_{7}$Se$_{8}$ & 26.8\cite{Kobayashi515} & 3.16\cite{powell014415} & 8.5
\\ 
&  &  &  \\ \hline
&  &  &  \\ 
Selenides &  &  &  \\ 
Fe$_{7}$Se$_{8}$ & 24.1\cite{Ok73} & 3.86 \cite{Andresen64} & 6.2 \\ 
&  &  &  \\ \hline
&  &  &  \\ 
Tellurides &  &  &  \\ 
Fe$_{1.125}$Te & --- & 2.07\cite{Fruchart169} &  \\ 
Fe$_{1+x}$Te & --- & 1.96--2.03\cite{Bao247001} &  \\ 
$0.076\leq x\leq 0.141$ &  &  &  \\ 
Fe$_{1.068}$Te & --- & 2.25\cite{Li054503} &  \\ 
Fe$_{1.05}$Te & --- & 2.54\cite{Martinelli094115} &  \\ 
Fe$_{1.11}$Te & 11\cite{hermon74} & --- &  \\ 
Fe$_{1.08}$Te & 10.34\cite{MizuguchiS338} &  & 4.3--5.2 \\ 
&  &  &  \\ \hline\hline
\end{tabular}%
\end{table}

Impurities may provide a possible origin for the variation in measured
moments. M\"{o}ssbauer spectroscopy, while sensitive to the presence of
impurity phases, does not rely on normalisation to the total sample in order
to determine moments, they come rather from the observed line splitting, and
not the intensity. Neutron diffraction, by contrast, while providing far
more information on the magnetic ordering, ultimately relies on peak
intensities, normalized to the total nuclear scattering, to determine the
magnetic moments. It is clear from the 10~K spectrum shown in Fig. \ref%
{Fig1.12} that there is a central paramagnetic component present that
involves about 12$\pm $2\% of the iron in the sample. Such high apparent
impurity levels in single crystal samples that had no impurities detected by
powder x-ray diffraction, deserves further attention. If the paramagnetic
component is not an \textquotedblleft impurity\textquotedblright\ then it
must either be intrinsic to the structure or a property of the material.

At the temperatures of interest here, K$_{0.8}$Fe$_{1.76}$Se$_{2}$ adopts a
vacancy-ordered $I4/m$ modification of the parent ThCr$_{2}$Si$_{2}\mathrm{-}
$type $I4/mmm$ structure with iron essentially filling a 16$i$ site and
leaving ordered vacancies on the (almost) empty 4$d$ site \cite{Pomj1919}$,$%
\cite{Zava}. Occupations of $\sim $8\% for the Fe-4$d$ site have been
reported \cite{Zava}. If we assume full occupation of the Fe-16$i$ site in
our sample, this leaves 9\% of the iron in the 4$d$ site. Partial occupation
of the Fe-16$i$ site would leave more iron to be accommodated in the 4$d$
site. As we see no evidence for a second magnetic component that could be
associated with iron in the 4$d$ site, it is possible that the iron in these
more isolated sites does not order, in which case our estimate of $%
\gtrapprox $9\% in the 4$d$ site is fully consistent with the 12$\pm $2\%
paramagnetic component observed in the M\"{o}ssbauer spectrum.

\begin{figure}[h]
\centerline{\includegraphics [scale=0.8]{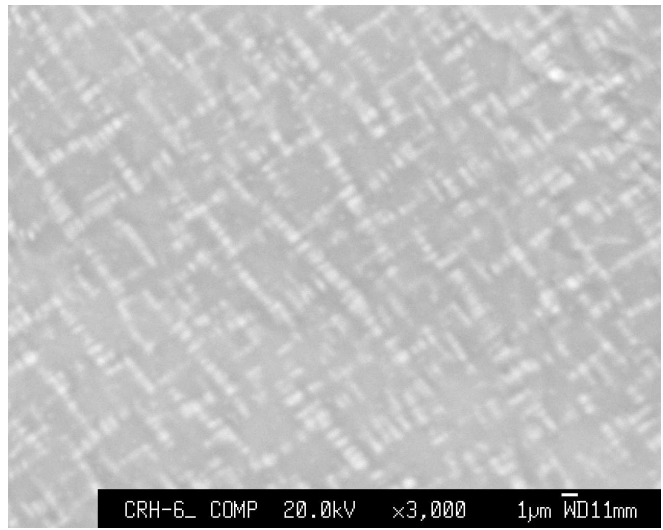}}
\caption{Backscattered electron analysis (BSE) image of a cleaved crystal
surface of $\mathrm{K_{0.80}Fe_{1.76}Se_{2.00}}$ taken at an accelerating
voltage of 20~kV. The lighter regions have lower potassium concentrations
than the darker background area. }
\label{Fig1.13}
\end{figure}

Another possible origin of the 12$\pm $2\% non-magnetic Fe component in the
low temperature (including room temperature) state can be seen in the
backscattered electron analysis (BSE) image shown in Fig. \ref{Fig1.13}.
This image reveals that there is, at the micron scale, a modulation in the
surface composition that can be correlated, through a preliminary line-scan
analysis of the WDS data, with reductions of K content or increase of Fe
content in the lighter regions. Auger electron analysis further confirmed
this observation and gave a rough estimate of a composition of K$_{0.9}$Fe$%
_{1.7}$Se$_{2}$ for the dark region and K$_{0.6}$Fe$_{1.9}$Se$_{2}$ for the
light region. It should be noted, though, that such patterns appear in
samples grown by furnace cooling as well as samples decanted from a liquid
melt\cite{Hu1931}.

\begin{figure}[h]
\centerline{\includegraphics [scale=0.8]{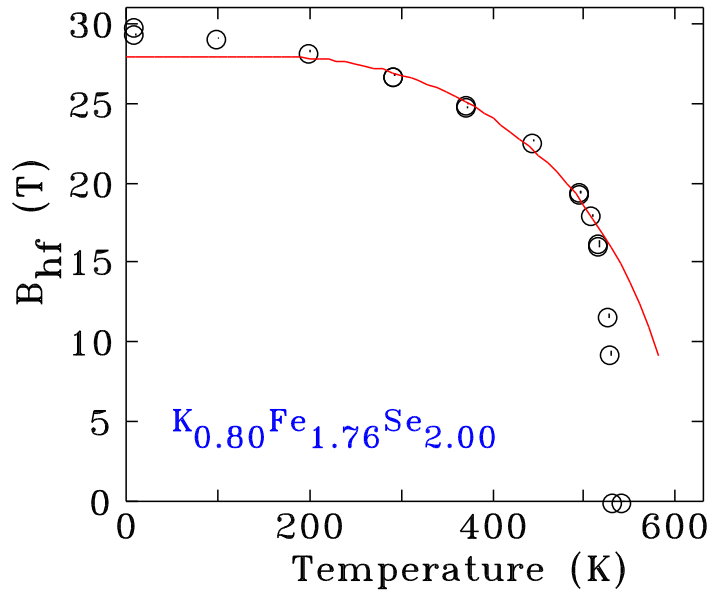}}
\caption{(color online) Temperature dependence of the magnetic hyperfine
field (B$_{hf}$) in $\mathrm{K_{0.80}Fe_{1.76}Se_{2.00}}$. The solid line is
a fit to a J=$\frac{1}{2}$ Brillouin function between 200~K and 500~K that
yields an expected transition of 600$\pm$30~K, well above the observed value
of 532$\pm$2~K. Fitted errors on B$_{hf}$ are less than 0.1~T, much smaller
than the plotting symbols. The rapid collapse above 500~K is accompanied by
the growth of a paramagnetic component (see Fig. \protect\ref{Fig1.15}). }
\label{Fig1.14}
\end{figure}

Raising the temperature leads to the expected decline in $B_{hf}$, however
it is clear from Fig. \ref{Fig1.12} that magnetic order persists up to
530~K, confirming that K$_{0.8}$Fe$_{1.76}$Se$_{2}$ has a remarkably high
ordering temperature. The temperature dependence of $B_{hf}$ shown in Fig. %
\ref{Fig1.14} yields an ordering temperature of $T_{N}=\,532\,\pm \,2$~K.
However this is not the result of the fit to a J=$\frac{1}{2}$ Brillouin
function shown in Fig. \ref{Fig1.14} as this predicts a transition
temperature of 600$\pm $30~K and the observed behaviour departs from this
curve above 500~K. The two points that bracket the transition are at 530~K,
where a clear magnetic signal is seen, and at 533~K where the sample is no
longer magnetic, setting the transition at $532\pm 2$~K.

A neutron diffraction study of K$_{0.8}$Fe$_{1.6}$Se$_{2}$ found two regions
in which the temperature dependence of the magnetic parameter was unusual 
\cite{Weibao}. From 50~K to 450~K they found a linear dependence of the
(101) magnetic peak intensity, suggesting that $\mu _{Fe}^{2}$ is a linear
function of temperature. The clear curvature of $B_{hf}$(T) in this region,
shown in Fig. \ref{Fig1.14}, is not consistent with this form, as squaring
our observed $B_{hf}$(T) to get something that would scale with the
scattering intensity in a neutron diffraction pattern leads to \textit{%
increased} curvature rather than linear behaviour.

\begin{figure}[h]
\centerline{\includegraphics [scale=0.8]{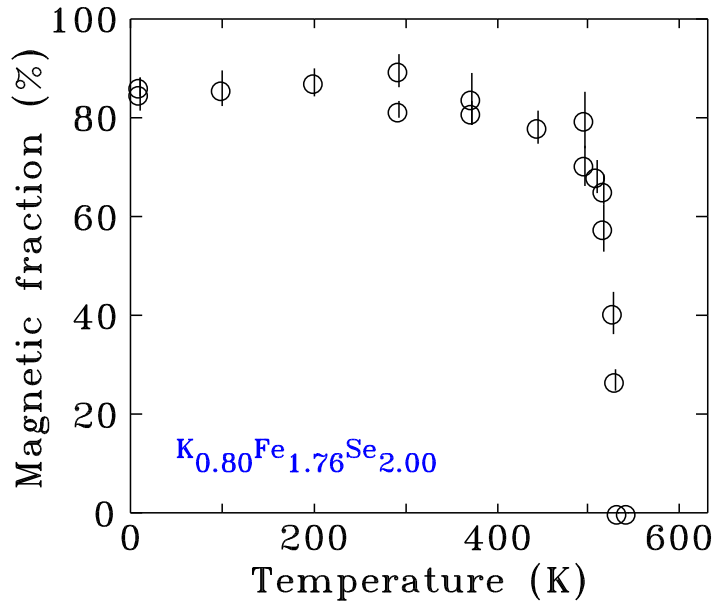}}
\caption{(color online) Temperature dependence of the magnetic fraction in $%
\mathrm{K_{0.80}Fe_{1.76}Se_{2.00}}$. The rapid collapse above 500~K
indicates that the magnetic transition has first order character and may be
associated with a structural transition. }
\label{Fig1.15}
\end{figure}

Above 500~K, Wei Bao \textit{et al.} reported a very rapid decrease in the
(101) intensity \cite{Weibao} leading to an ordering temperature of $\sim $%
560~K. Whereas our sample composition is slightly different and our ordering
temperature slightly lower, we see the \textit{same} abrupt loss of magnetic
order in Fig. \ref{Fig1.14}. It is noteworthy that magnetic susceptibility
measurements show somewhat smaller $T_{N}$ than that revealed by neutron
diffraction but very similar $T_{N}$ to that indicated by M\"{o}ssbauer
spectrum in this article.\cite{Weibao2}$,$\cite{YJYan}$,$\cite{RHLiu}
Inspection of the spectra above 500~K shown in Fig. \ref{Fig1.12} reveals
that the intensity of the magnetic peaks decreases visibly as their
splitting falls. The ability to uniquely separate the amount of a magnetic
phase (seen through line intensities) from the magnitude of the magnetic
order (seen independently through line splittings) is an important strength
of M\"{o}ssbauer spectroscopy. Tracking the fraction of the iron that is
present as a magnetically ordered form (Fig. \ref{Fig1.15}) confirms that
the magnetic phase is disappearing even faster than the splitting that marks
the order. This strongly suggests that the magnetic phase is transforming
before it reaches its true ordering temperature (which we estimate to be
about 600~K) and that the observed transition is being driven by a first
order structural event. This view is supported by the neutron diffraction
work of Wei Bao \textit{et al.} \cite{Weibao} where they also tracked the
intensity of the (110) structural peak that is associated with the $I4/m$
vacancy-ordered structure of b below 580~K. This peak starts to lose
intensity at the same temperature at which the (101) magnetic peak starts
its sudden decline. As we see both a weakening of the magnetic order and a
reduction in the magnetic fraction above 500~K, it is possible that the
break-up of the vacancy-ordered magnetic form reduces the magnetic
connectivity of the ordered phase until it forms a non-percolating network
of finite clusters. The magnetic order is then lost at a temperature below
both its intrinsic ordering temperature, and the temperature at which the
vacancy-ordered $I4/m$ structure fully transforms to the high-temperature $%
I4/mmm$ form.\newline

\subsection{Phase separation and possible superconducting aerogel}

The data presented so far offer a rather contradictory set of observations.
On one hand K$_{0.8}$Fe$_{1.76}$Se$_{2}$ appears to have a high value of $%
T_{c}$, a fair-sized shielding fraction, and $H_{c2}(T)$ anisotropy that is
consistent with many of the other Fe-based superconductors. On the other
hand, the electrical resistivity of K$_{0.8}$Fe$_{1.76}$Se$_{2}$ is
anomalously high, with increasingly non-metallic temperature dependence
depending on precise Fe stoichiometry, the specific heat jump, $\Delta
C_{p}/T_{c}$ is relatively small, and there is large, local moment-like
antiferromagnetic order of the Fe sublattice with a first order transition
near 530 K. If the sample were to be considered homogeneous, with all of
these features being associated with the same, single phase, then we would
need to consider K$_{0.8}$Fe$_{1.76}$Se$_{2}$ to be an anomalous and very
different type of Fe-based superconductivity.

As it stands, though, there are several indications, in the data presented,
that K$_{0.8}$Fe$_{1.76}$Se$_{2}$ is not homogeneous, but rather is phase
separated into a non-magnetically ordered, minority phase that is
superconducting and a majority phase that manifest high temperature, large
moment, antiferromagnetic order and is probably near insulating.

The M\"{o}ssbauer and electron microscopy data shown in Figures 1.12 - 1.15
indicate that there may well be a mesoscopic phase separation into a
majority phase with a large hyperfine field on the Fe site and a minority
phase with essentially no hyperfine field on the Fe site. This, combined
with the reduced jump in $\Delta C_{p}/T_{c}$ and the apparently high
electrical resistivity, point toward a scaffold-like network (or
aerogel-like pattern) of conducting (and below 30 K, superconducting) phase
that exists on a sub-micron length scale. Such a network would be consistent
with the moderate shielding seen in the ZFC low field magnetization data as
well as the magneto optical results. Such a phase separated, minority,
superconducting phase would allow the majority phase to host the large
moment antiferromagnetism and also be a poor conductor, explaining the
curious composite resistivity data that is so sensitive on Fe stoichiometry.
It is important to note, though, that given the clear anisotropy in $%
H_{c2}(T)$ data, this minority phase has to remain, at least partially,
oriented with the host matrix. Given the similarity in $H_{c2}(T)$
anisotropy between K$_{0.8}$Fe$_{1.76}$Se$_{2}$ and other Fe-based
superconductors, it is tempting to assume that the orientation of the
minority phase is closer to complete than to random, but this will need to
be tested directly. After our initial reports of phase separation\cite%
{Ryan2011}, the phase separation scenario in K$_{x}$Fe$_{2-y}$Se$_{2}$ is
proposed by more and more recent reports, from Transmission Electron
Microscopy\cite{JQLi},\cite{ZWang}, single-crystal XRD\cite{Ricci} and
magnetic hysteresis loops.\cite{HHWenEPL}

\section{Summary}

K$_{x}$Fe$_{2-y}$Se$_{2}$ superconductor has been extensively studied since
its discovery. But the non-stoichiometry present a complex material problem
that hinders better understanding of its properties. Our single crystal
growth from the furnace-cooled and decanted methods implies that it is grown
from excess Fe-Se flux and crystals with similar stoichiometry can be
obtained from both methods with $T_{c}$ $\sim $ 30 K. Single crystals of K$%
_{0.8}$Fe$_{1.76}$Se$_{2}$ exhibit moderate anisotropy in both magnetic
susceptibility and electrical resistivity with $\chi _{ab}/\chi _{c}\sim 2$
and $\rho _{c}/\rho _{ab}\sim 4$ at 300 K. Broadened superconducting
transitions seen in several measurements may be associated with a small
variation of stoichiometry of the crystal, consistent with what was shown by
WDS analysis. The upper critical field of K$_{0.8}$Fe$_{1.76}$Se$_{2}$ is
determined as $H_{c2}^{ab}$(18 K) $\simeq $ 54 T and $H_{c2}^{c}$(1.6 K) $%
\simeq $ 56 T. The anisotropy parameter $\gamma _{H}$ initially increases
with decreasing temperature, passed through a maximum of $\sim 3.6$ near 27
K, then decreases to $\sim 2.5$ at 18 K. The observed $\gamma _{H}$ values
show a weakening anisotropic effect at low temperatures. Although the
Fe-based superconductors have a layered crystal structure, a weak anisotropy
of $H_{c2}$ may be a common feature, suggesting that the inter-layer
coupling and the three dimensional Fermi surface may play an important role
in the superconductivity of this family. Our $^{57}$Fe M\"{o}ssbauer
spectroscopy study confirms the presence of magnetic order from well below $%
T_{c}\sim 30$~K to $T_{N}=\,532\,\pm \,2$~K. The large magnetic splitting of
29.4$\pm $0.1~T at 10~K indicates that the iron moments are large,
consistent with values of 3.3~$\mu _{B}$/Fe observed by neutron diffraction
at 11~K\cite{Weibao}, while the line intensities indicate that the ordering
is almost parallel to the \textit{c}-axis. An apparent paramagnetic impurity
phase can be attributed to iron atoms in the 4$d$ site or the stoichiometry
variation of the microstructure seen in BES image. Analysis of the spectra
taken in the vicinity of $T_{N}$ shows that the magnetic fraction decreases
rapidly above 500~K and that the loss of order is driven by a first order
structural transition.

In addition to the above, conspicuous observations, there is growing
evidence that K$_{0.8}$Fe$_{1.76}$Se$_{2}$ is a phase separated sample, with
a metallic (and at low temperature, superconducting) minority phase that
does not manifest long range magnetic order and a majority phase that
undergoes a first order, antiferromagnetic phase transition near 530 K and
may well be either non-conducting or very poorly conducting. As such, K$%
_{0.8}$Fe$_{1.76}$Se$_{2}$ would essentially be a superconducting aerogel
embedded in a matrix of antiferromagnetic (near) insulator. If there is
indeed such a phase separation, then K$_{0.8}$Fe$_{1.76}$Se$_{2}$ can be
understood, or at least categorized, as another example of Fe-based
superconductivity similar at a qualitative level to other, better
understood, and single phase, examples.

\section*{Acknowledgements}

This work was carried out at the Iowa State University and supported by the
AFOSR-MURI grant \#FA9550-09-1-0603 (R.H. and P.C.C.). Part of this work was
performed at Ames Laboratory, US DOE, under contract \# DE-AC02-07CH 11358
(K.C., H.K., H.H., W.E.S., M.A.T., R.P., S.L.B. and P.C.C.). S.L.B. also
acknowledges partial support from the State of Iowa through Iowa State
University. R.P. acknowledges support from the Alfred P. Sloan
Foundation.Work at the NHMFL-PFF is supported by the NSF, the DOE and the
State of Florida. M\"{o}ssbauer work was supported by the Natural Sciences
and Engineering Research Council of Canada and Fonds Qu\'{e}b\'{e}cois de la
Recherche sur la Nature et les Technologies. J.M.C. acknowledges support
from the Canada Research Chairs programme.

\end{document}